\documentclass[grl,draft]{agutex}
\usepackage{graphicx}
\usepackage{color}
\usepackage{ulem}

\authorrunninghead{Lamy et al.}
\titlerunninghead{Properties of SKR within its source region}

\begin{document}

\title{Properties of Saturn Kilometric Radiation measured within its source region}

\authors{L. Lamy\altaffilmark{1}, P. Schippers\altaffilmark{2}, P. Zarka\altaffilmark{3}, B. Cecconi\altaffilmark{3}, C. Arridge\altaffilmark{4}, M.K. Dougherty\altaffilmark{1}, P. Louarn\altaffilmark{5}, N. Andr\'e\altaffilmark{5}, W.S. Kurth\altaffilmark{2}, R.L. Mutel \altaffilmark{2}, D.A. Gurnett \altaffilmark{2}, A.J. Coates\altaffilmark{4}}

\altaffiltext{1}{Space and Atmospheric Physics, Blackett Laboratory, Imperial College London, London, UK}
\altaffiltext{2}{Department of Physics and Astronomy, University of Iowa, Iowa City, Iowa, USA}
\altaffiltext{3}{Laboratoire d'Etudes et d'Instrumentation en Astrophysique, Observatoire de Paris, CNRS, Meudon, France}
\altaffiltext{4}{Centre d'Etude Spatiale des Rayonnements, Universit\'e Paul Sabatier, CNRS, Toulouse, France}
\altaffiltext{5}{The Centre for Planetary Sciences at UCL/Birkbeck, London, UK}

\begin{abstract}
On 17 October 2008, the Cassini spacecraft crossed the southern sources of Saturn kilometric radiation (SKR), while flying along high-latitude nightside magnetic field lines. In situ measurements allowed us to characterize for the first time the source region of an extra-terrestrial auroral radio emission. Using radio, magnetic field and particle observations, we show that SKR sources are surrounded by a hot tenuous plasma, in a region of upward field-aligned currents. Magnetic field lines supporting radio sources map a continuous, high-latitude and spiral-shaped auroral oval observed on the dawnside, consistent with enhanced auroral activity. Investigating the Cyclotron Maser Instability (CMI) as a mechanism responsible for SKR generation, we find that observed cutoff frequencies are consistent with radio waves amplified perpendicular to the magnetic field by hot (6 to 9~keV) resonant electrons, measured locally.
\end{abstract}

\begin{article}

\section{Introduction}

In the past four decades, remote observations have identified intense radio emissions from auroral regions of all explored magnetized planets of the solar system \citep{zarka_JGR_98}. However, the source region of the terrestrial auroral kilometric radiation (AKR) has been the only one studied so far with in situ measurements. First observations from ISIS 1 (1970's), followed by extensive ones from Viking (1980's) and FAST (1990's) brought crucial constraints on the local plasma conditions and properties of emitted waves, leading to a comprehensive picture of AKR and its generation mechanism.

AKR is a cyclotron emission mainly emitted on the extraordinary (X) mode by auroral electrons, at frequencies close to the local gyrofrequency $f_\mathit{ce}$, between 50 and 700~kHz, in high-latitude tenuous regions (or auroral cavities). Signatures of AKR source crossings were identified in Viking dynamic spectra above 100~kHz by (i) an enhanced amplitude of the low-frequency emissions, typically reaching 10~mV.m$^{-1}$, and (ii) a cutoff frequency $f_\mathit{cut}$ close to, and occasionally below, $f_\mathit{ce}$ \citep{bahnsen_JGR_89, roux_JGR_93}. X mode emission at $f\le f_\mathit{ce}$ was related to predominant weakly relativistic (typically 5~keV) electrons \citep{lequeau_JGR_89}. Indeed, radio sources lie in auroral cavities depleted in cold plasma ($\le1$~keV), where $f_\mathit{pe}/f_\mathit{ce}\le0.1$, and 5 to 10 times less dense than the surrounding medium. These cavities correspond to acceleration regions with large electric potential drops, characterized by downward (upward) beams of electrons (ions) \citep{benson_GRL_79}, along field lines connected to auroral arcs. The spatial extent of AKR source regions is elongated in altitude (0.5 to 3~R$_\mathit{E}$ according to the AKR spectrum, 1~R$_\mathit{E}=6378$~km), narrow in latitude (a few tens of km), more extended in longitude and time variable \citep{hilgers_GRL_91}. Attributed to the wave-electron interaction named Cyclotron Maser Instability (CMI) \citep{wu_ApJ_79}, AKR is emitted quasi-perpendicularly from the local magnetic field \citep{hilgers_JGR_92} by unstable trapped \citep{louarn_JGR_90}, or shell-type \citep{ergun_ApJ_00} electron distributions.

Saturn Kilometric Radiation (SKR) is the kronian equivalent of AKR. Remote observations by Voyager (flybys in 1980 and 1981) and Cassini (in orbit since mid-2004) have established that SKR is also mainly observed on the X mode between 3 and 1200~kHz ($i.e.$ altitudes between 0.1 and 5~R$_\mathit{S}$, 1~R$_\mathit{S}=60268$~km), consistent with CMI, along magnetic field lines globally associated with atmospheric aurorae \citep{kurth_09}. However, in the absence of in situ measurements, the characteristics of the SKR source region, together with the nature and origin of the unstable resonant electrons responsible for its generation (or source of free energy), remained unknown.

The first crossing of the SKR source region was identified by Cassini on day 291 of 2008 near midnight Local Time (LT), at a distance of 5~R$_S$ ($f_\mathit{ce}\sim10$~kHz) and latitudes below $-60^\circ$ \citep{kurth_GRL_10}. Here, we investigate simultaneous observations of the Radio and Plasma Wave Science (RPWS) experiment, the magnetometer (MAG) and the Cassini Plasma Electron Spectrometer (CAPS) instrument (see auxiliary material for details of the specific data processing). We focus on the characteristics of the source region (section \ref{source_region}),  the locus of local and distant radio sources (section \ref{location}), and investigate how the CMI mechanism can lead to the observed SKR frequencies (section \ref{CMI}). The results are compared with the terrestrial case.

\section{Source region properties}
\label{source_region}

RPWS, MAG and CAPS observations, for day 291 of year 2008 are displayed in Figure \ref{fig1} between 0600 and 1100~UT. Figure \ref{fig1}a shows a dynamic spectrum of electric spectral density measured on one RPWS monopole between 3.5 and 1500~kHz. Except around 0848 and 0924~UT, where the antenna detected little signal above 100~kHz, due to its quasi-alignment with incoming wave vectors, the SKR spectrum extends from $\sim f_\mathit{ce}$ to 1000~kHz. Criteria (i) and (ii), as used to identify AKR source crossings at Earth, are clearly satisfied at low frequencies, around 10~kHz. First, (i) the low-frequency SKR amplitude particularly increases. Indeed, once calibrated, corrected for the actual distance to each source (see section \ref{location}) and normalized to 1~AU, the SKR intensity reaches $\sim$~10$^{-18}$ W.m$^{-2}$.Hz$^{-1}$, $i.e.$ the upper 1\% quantile of SKR mid-latitudes observations at 10~kHz \citep{lamy_JGR_08}. Then, (ii) the SKR spectrum shows a cutoff around $f_\mathit{ce}$. Frequencies $f_\mathit{cut}$ (see auxiliary text and Figures S1,2), as well as $f_\mathit{ce}$, are quantitatively investigated in Figure \ref{fig1}b. Whereas $f_\mathit{cut}$ approaches $f_\mathit{ce}$ from 0748 to 1000~UT, the amplitude is especially intense close to $f_\mathit{ce}$ between 0812 and 0912~UT (orange shaded, hereafter called the source region), with three unambiguous excursions below $f_\mathit{ce}$ (red shaded, labelled A, B and C). There, the SKR peaks at $\sim$~10$^{-9}$ V$^2$.Hz$^{-1}$ within the [$f_\mathit{ce}-$1kHz,$f_\mathit{ce}$+4kHz] bandwidth, corresponding to a field strength of $\sim$~1~mV.m$^{-1}$. Events A, B and C display negative $(f_\mathit{cut}-f_\mathit{ce})/f_\mathit{ce}$, as low as $\sim-2\%$, considered at Earth as the best indication of AKR sources, as a result of predominant hot electrons. This feature is investigated in section \ref{CMI}.

Frequencies $f_\mathit{cut}$ and $f_\mathit{ce}$ are reproduced in the logarithmic plot of Figure \ref{fig1}c, together with the plasma frequency $f_\mathit{pe}$ (density) and its contributions from cold, warm, and hot electrons, derived from CAPS measurements. The uncertainty on the above values (typically $\le30\%$, see auxiliary Figure S3) is mainly due to the underestimate of cold electrons masked by the spacecraft positive potential, and electrons missed by incomplete and time variable pitch angle coverage. The total density does not display any terrestrial-like cavity devoid of cold electrons (where $f\le f_\mathit{ce}$) and surrounded by a denser and colder medium (where $f\ge f_\mathit{ce}$). Rather, we observe a large tenuous region, comprising a sporadic warm component, and dominated by hot electrons after 0700~UT, for latitudes $\lambda_\mathit{sc}\le-50^\circ$ and distances $4.1\le r_\mathit{sc}\le5.8$~R$_\mathit{S}$. Corresponding relative altitudes being much larger than at Earth, possible auroral cavities may be present closer to Saturn. Within the source region, the total electron plasma frequency $f_\mathit{pe}$ varies between 400 and 900~Hz, leading to $f_\mathit{pe}/f_\mathit{ce}$ ratios of 0.05-0.09, in agreement with requirements for CMI resonance and observed AKR generation conditions. Contrary to the situation at Earth, the kronian auroral plasma environment appears to be naturally tenuous/magnetized enough to generate SKR, where CMI-unstable electrons are present.

Figure \ref{fig1}d investigates the azimuthal component of the magnetic field. Gradients in $B_{\phi}$ indicate field-aligned currents (FAC) where the sign of the slope gives the direction (up- or downgoing) of the local net current. The source region (orange) corresponds to a positive slope starting at 0748~UT, and events A, B and C nearly match the steepest positive slopes. This trend suggests a globally upward current, associated with downward electrons, and consistent with observed upward ions \citep{kurth_GRL_10}, similar to observations within AKR sources. We also note a strong positive slope around 0630~UT, where no signature of SKR sources was detected. This can be explained by $f_\mathit{pe}/f_\mathit{ce}\ge0.5$ preventing CMI resonance.

MAG measurements thus complement CAPS observations shown by the electron spectrogram of Figure \ref{fig1}e, where the incomplete pitch angle coverage did not allow the detection of down-going electrons with pitch angles $\le15^\circ$. The phase space density of detected electrons reveals two sporadic distributions: a partial shell-like distribution for 1-10~keV (hot) electrons, that displays part of the expected down-going electrons, most likely accelerated at higher altitudes, and up-going field-aligned beams of a few 100~eV (warm) electrons. Both distributions, possible candidates for SKR generation, are observed during and outside of events A, B and C. They might thus complete the criteria used to identify source regions. Finally, no loss cone was observed in the electron distributions. The predicted loss cone angle $\alpha\sim5^\circ$ at such altitudes, was however smaller than the CAPS angular resolution.

\section{Location of SKR sources}
\label{location}

RPWS-HFR 3-antenna observations unambiguously provide the wave vector {\bf k} of each time-frequency measurement, assuming transverse waves and instantaneous point sources \citep{cecconi_RS_05}. Under the hypothesis of SKR generation at $f=f_\mathit{ce}$, \citet{cecconi_JGR_09} used the SPV magnetic field model, and a simple current sheet, to derive the intersection of each {\bf k} vector with the corresponding isosurface $f=f_\mathit{ce}$, giving the 3D spatial source location and its associated field line.

This remote sensing capability was used within the source region to determine the instantaneous 3D location of both local and distant radio sources. Figures \ref{fig3}a,b show the spatial distribution of low-frequency SKR sources over the interval [0812,0912]~UT, as observed from Cassini and after magnetic polar projection. The radio source localization technique yields more accurate locations for sources closer to the spacecraft, $i.e.$ lowest frequencies. Radio emissions at $f\ge f_\mathit{ce}$+1kHz reveal distant sources (blue) organized along dawnside high-latitude magnetic field lines, with, as expected, sources closer to the planet for increasing frequencies. Footprints of associated field lines, known to map to part of a circumpolar radio auroral oval \citep{lamy_JGR_09}, are identified in the polar view to vary from latitudes $\lambda_B=-80\pm2^\circ$ at LT$_B=0100$ to $-75\pm5^\circ$ at 0700. 

Low-frequency emissions $f\le f_\mathit{ce}+1$~kHz (red) appear as local sources, widely distributed in all directions in Figure \ref{fig3}a, while strikingly matching the ionospheric footprint of the field lines crossed by Cassini on the polar view (green line). Based on the spacecraft velocity, the dimension of the overall source region along Cassini's trajectory is $\sim51000$~km. Events A, B, and C, previously identified as traversed sources, display shorter and similar dimensions of 1800~km for event A, and 900~km for events B and C, corresponding to latitudinal widths of $\sim0.2^\circ$ and $\sim0.1^\circ$ respectively, comparable to $\sim0.2^\circ$ values at Earth (for a typical AKR source at a 3~R$_E$ altitude and a north-south extent of 100~km), but with sources located much farther from the planet. This suggests successive ($\ge3$) encounters of SKR curtains, consistent with sub-ovals often observed in UV aurorae [W. Pryor, pers. com.], even if the accuracy of the radio source location was not sufficient to confirm it here. 

The unusual distribution of local and distant sources in Figure \ref{fig3}b is confirmed and expanded along an extended time-frequency interval by the radio image of Figure \ref{fig3}c, that maps SKR intensity onto the polar region. It reveals half of the entire radio auroral oval, with the most intense emissions corresponding to the lowest frequencies within the source region shown by Figures \ref{fig1}a and \ref{fig3}a,b. This distribution also lies along an extended spiral shape, whose latitude, and latitudinal extent, decreases, and broadens, with LT until $\lambda_B=-75\pm6^\circ$ at LT$_B=1300$.

The crossing of intense SKR sources at low frequencies (high altitudes) located close to midnight, coincided with field lines characterizing a spiral auroral oval, starting from very high latitudes on the nightside, and filling in the dawn sector of the polar cap. This unusual distribution of active magnetic field lines is reminiscent of the well known active UV aurorae, possibly triggered by a solar wind compression \citep{bunce_JGR_10}, and followed by a nightside injection related to plasmoid activity \citep{jackman_JGR_09}.

\section{CMI resonance and SKR emission frequency}
\label{CMI}

In regions where $f_\mathit{pe}/f_\mathit{ce}$ is lower than $\sim0.1$ and unstable electron distributions are present, CMI emission can be generated by electrons that fulfill the resonance equation: $f=f_\mathit{ce}/\Gamma+k_{\parallel}v_{\parallel}$, where {\bf k} is the wave vector, {\bf v} the electron velocity, the subscript ${\parallel}$ refers to the direction parallel to the magnetic field, and $\Gamma=1/\sqrt{1-v^2/c^2}$ is the Lorentz factor \citep{wu_ApJ_79}. At Earth, AKR being mostly emitted perpendicular to the magnetic field, this equation reduces to $f=f_\mathit{ce}/\Gamma$ . The CMI emission frequency thus lies below $f_\mathit{ce}$, the difference being directly determined by the kinetic energy of resonant electrons.

Assuming nearly perpendicular emission of SKR, Figure \ref{fig2} shows that hot electrons are the only reliable candidate for SKR generation. Indeed, their kinetic energy, between 6 and 9~keV, leads to $f$ (red) $\sim2\%$ below $f_\mathit{ce}$ (blue), that match observed $f_\mathit{cut}$ (black) during events A, B and C within error bars. Precisely, the correlation coefficient computed over these 3 events (8 measurements) between $f$ and $f_\mathit{cut}$ reaches $c~=~0.94$. Warm (and cold) electrons lead to frequencies that cannot be distinguished from $f_\mathit{ce}$ ($\Gamma\sim$1), $i.e.$ inconsistent with the observed $f_\mathit{cut}$.

In the case of non-perpendicular emission of SKR, the resonance frequency is modified by the contribution of $k_{\parallel}v_{\parallel}$. Unstable electron distributions, such as loss cone or ring distributions, can drive CMI oblique emission from upgoing electrons. Since radio waves can only be radiated away from the planet, this leads to $k_{\parallel}v_{\parallel}\ge0$, which, in turn, implies CMI frequencies higher than the ones derived for purely perpendicular SKR. More precisely, $k_{\parallel}=N\omega/c \cos\theta$, with N the index of refraction, $\theta$ the angle of emission with respect to the local magnetic field vector, and $v_{\parallel}=v\cos\alpha$, with $\alpha$ the loss cone angle. Using $\alpha\sim5^\circ$, $\theta\sim70^\circ$ (compatible with RPWS measurements), and assuming N~$\sim1$, we computed the contribution of $k_{\parallel}v_{\parallel}$, and therefore CMI emission frequencies for cold, warm and hot electrons. None of these frequencies lie below $f_\mathit{ce}$ and thus cannot account for the observed $f_\mathit{cut}$.

In summary, we have shown that the SKR cutoff frequencies observed below $f_\mathit{ce}$ can only be explained by a perpendicular SKR emission, excited by the 6 to 9~keV (hot) electrons measured by CAPS. In addition, we note that the shell-like distribution of hot electrons mentioned in section \ref{source_region} is expected to lead to perpendicular emission \citep{ergun_ApJ_00}. These results are consistent with the terrestrial case.

\section{Conclusion}

Cassini crossed the first source region of exo-terrestrial radio auroral emissions, evidenced by (i) intense SKR at low frequencies, and (ii) $f_\mathit{cut}\le f_\mathit{ce}$. 

Among the similarities with AKR at Earth, the SKR source region at Saturn is characterized by $f_\mathit{pe}/f_\mathit{ce}\le0.1$, and upward field-aligned currents, consistent with electrons accelerated toward the planet. Local sources, emitted at frequencies below, or close to, $f_\mathit{ce}$, are detected on field lines crossed by Cassini. The auroral source region is dominated by hot electrons, found to be the only reliable source of energy for CMI resonance, and implying SKR emission perpendicular to the magnetic field.

In contrast to Earth, the SKR sources were detected at much lower frequencies (10~kHz), $i.e.$ much higher altitudes (4.1~R$_\mathit{S}$). The auroral region does not reveal any terrestrial-like cavity, devoid of cold plasma, but appears to be naturally tenuous/magnetized enough to permit CMI driven emission, where unstable electrons are present.

This study revealed intense sources crossed at unusual locations (nightside, very high latitude), and part of an extended spiral radio auroral oval, with high latitude dawnside emission, that suggests enhanced auroral activity. More crossings of the SKR source region, planned at the end of the Cassini mission, are needed to study statistically those characteristics, possibly varying with location and time.

\begin{acknowledgments}
We thank Cassini RPWS, MAG and CAPS engineers for support on instrumental questions. The French co-authors acknowledge support from CNES agency. LL thanks A. Roux, M. Dekkali, R. Prang\'e, F. Mottez and S. Hess for inspiring discussions. LL and MKD were supported by the STFC rolling grant to ICL. CSA was supported by an STFC Postdoctoral fellowship and AJC by the STFC rolling grant to MSSL/UCL. The research at The University of Iowa is supported by NASA through Contract 1356500 with the JPL. 
\end{acknowledgments}


\begin{figure*}
\centering
\includegraphics[width=40pc]{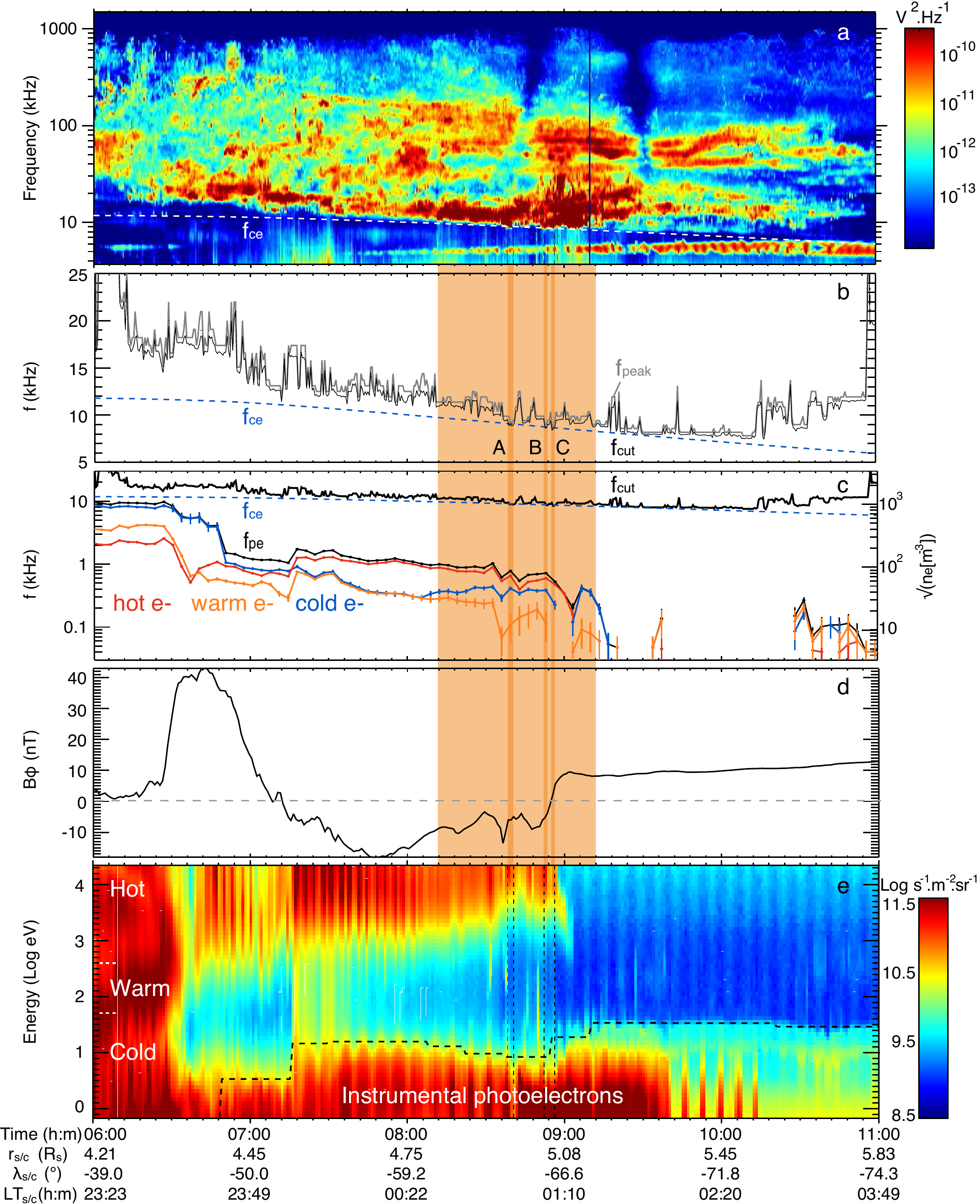}
\end{figure*}

\begin{figure*}
\centering
\caption{Multi-instrument observations during day 291 of year 2008 over [0600,1100] UT. Panel a displays the RPWS-HFR dynamic spectrum of electric spectral density recorded by the Z monopole (rather than the Stokes parameter S, computed from simultaneous 2-antenna measurements, but noisier at low frequencies). The dashed line indicates the local $f_\mathit{ce}$, as derived from MAG observations. SKR lies between $\sim f_\mathit{ce}$ and 1000~kHz. Panel b superimposes $f_\mathit{peak}$ (gray) and $f_\mathit{cut}$ (black) to $f_\mathit{ce}$ (dashed), with a time resolution of 32~s. The typical uncertainty is less than 0.5~\% for $f_\mathit{cut}$, and less than 0.05~\% (resp. 0.5~\%) for $f_\mathit{ce}$ after (resp. before) 0821~UT. Orange shaded region maps the interval [0812,0912]~UT, where SKR is enhanced below 30~kHz and its low-frequency cutoff reaches $f_\mathit{ce}$. Red shaded events A, B and C mark unambiguous excursions $f_\mathit{cut}\le f_\mathit{ce}$. Panel c reproduces $f_\mathit{cut}$ and $f_\mathit{ce}$ with a log scale, together with the total electron density $n_e$ (expressed in $\sqrt{n_e(cm^{-3})}$, right, and $f_\mathit{pe}$(kHz), left) derived from CAPS-ELS observations (black dots). Colors reveal contributions of cold (blue, $\le50$~eV), warm (orange, 50-400~eV) and hot (red, $\ge400$~eV) electrons. Panel d shows the azimuthal magnetic field component B$_\phi$, measured by MAG with a a 1min resolution and a 0.1nT uncertainty. As a consequence of Maxwell-Amp\`ere equation, its variations indicate field-aligned currents.  A positive (negative) slope corresponds to an up-going (down-going) net current. Panel e displays the CAPS-ELS electron spectrogram of differential energy flux (DEF), summed over all anodes at a 3min resolution. Three electron populations are detected: a hot component above 400eV, a cold one below 50eV and a sporadic warm one in between. Cold electrons are partially masked by spacecraft photoelectrons, and all field-aligned down-going electrons were missed due to incomplete pitch angle coverage.}
\label{fig1}
\end{figure*}


\begin{figure*}
\noindent\includegraphics[width=39pc]{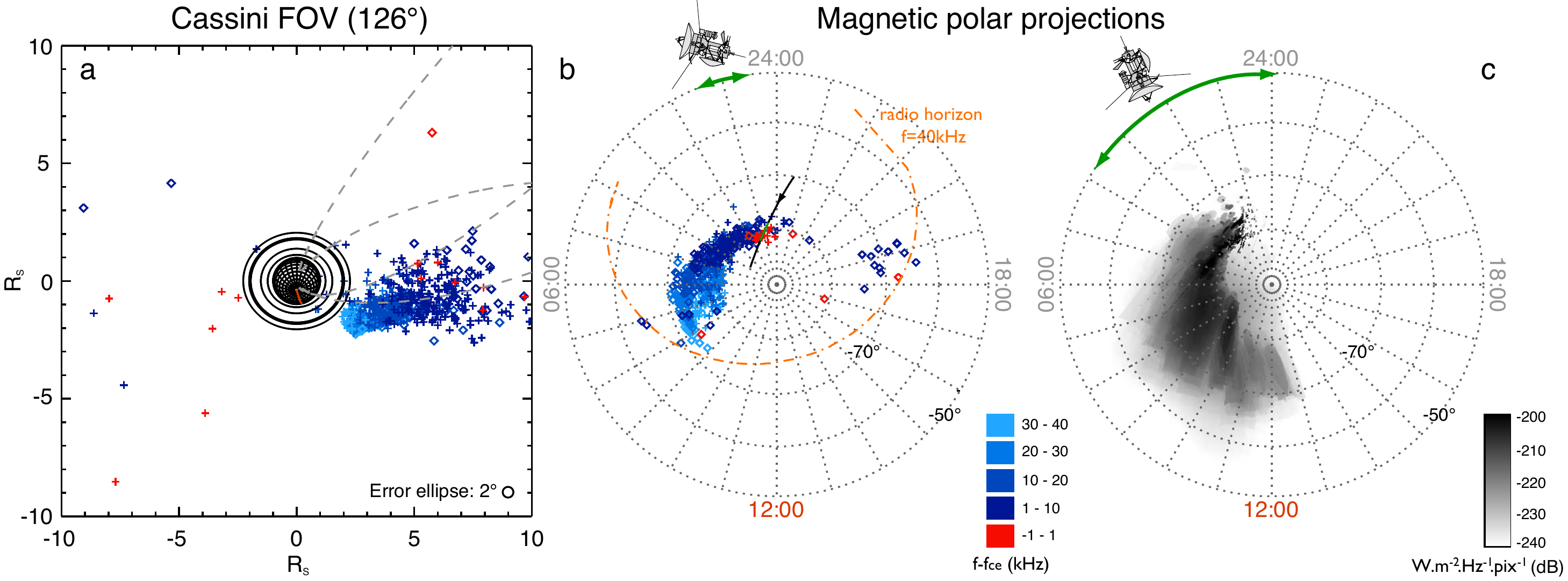}
\caption{Location of radio sources in Cassini's field of view, or FOV, (panel a) and magnetically projected down to the planet (panel b). The radio data selection is described in the auxiliary material. The time-frequency interval is restricted to [0812,0912]~UT (source region) and [7,40kHz] (highest resolution). Crosses (diamonds) refer to directions of arrival that cross (do not cross) their associated isosurface-$f=f_\mathit{ce}$ \citep{cecconi_JGR_09}. Most of the frequencies (plotted in blue scale for increasing ranges of $f-f_\mathit{ce}$) correspond to distant sources, clustered along high-latitude magnetic field lines in panel a (dashed, with footprint coordinates LT$_B$=[0400,0600] and $\lambda_B$=[-80$^\circ$,-75$^\circ$]), quantitatively identified on the dawnside in panel b. Frequencies plotted in red ($f\le f_\mathit{ce}$+1kHz) reveal local sources, detected through the entire observation plane in panel a, and mapped the ionosphere along field lines crossed by Cassini in panel b, the green (black) line showing the projected spacecraft trajectory for [0812,0912] UT ([0600,1100] UT). Panel c displays the radio map \citep{lamy_JGR_09} integrated along the extended time interval [0700,1100]~UT and frequency range [7,1000kHz]. The global distribution of the footprints of field lines supporting SKR sources follows an unusual spiral shape, starting from narrow very high latitude  $\sim$-80$\pm2^\circ$ at LT$_B$=0100, then evolving to broad lower latitudes toward noon until $\lambda_B$=-75$\pm$6$^\circ$ at LT$_B$=1300. }
\label{fig3}
\end{figure*}


\begin{figure*}
\centering
\noindent\includegraphics[width=40pc]{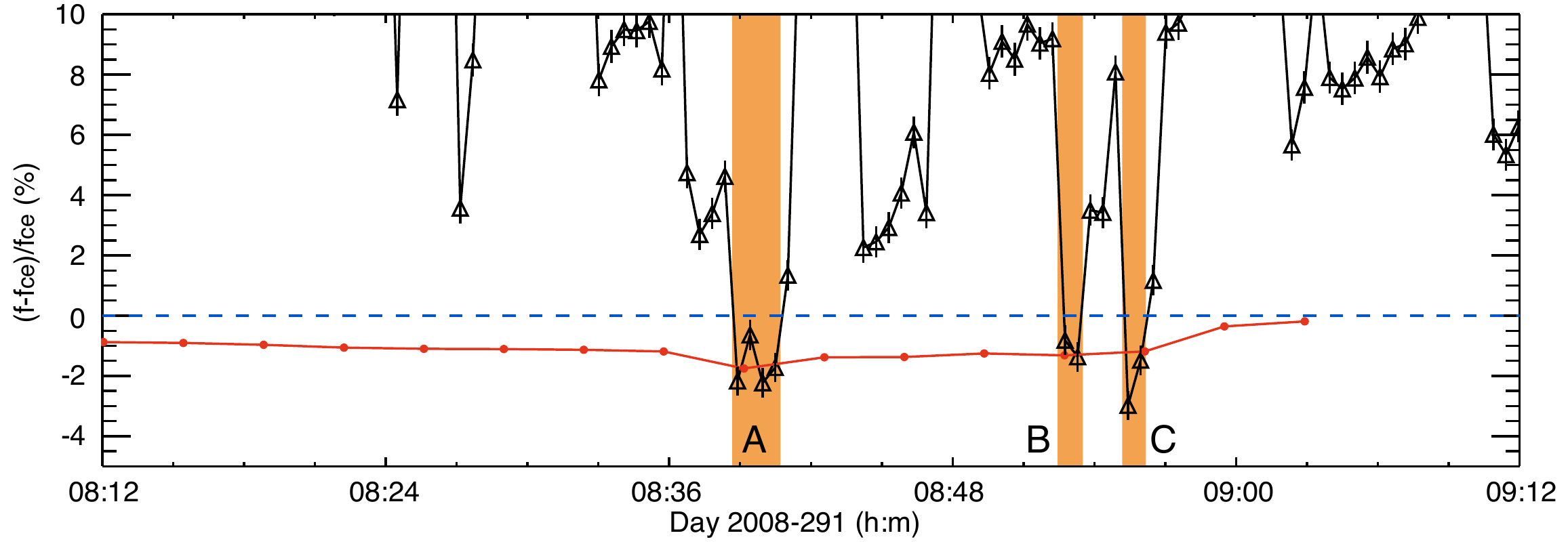}
\caption{Characteristic frequencies between 0812 and 0912~UT of day 2008-291, expressed relatively to $f_\mathit{ce}$. The SKR cutoff frequency f$_\mathit{cut}$, determined with an accuracy of 0.5~\%, is displayed in black, and f$_\mathit{ce}$, in dashed blue. The red curve refers to the CMI resonance frequency $f_\mathit{ce}/\Gamma$ computed with hot electrons observed by CAPS under the assumption of perpendicular SKR emission. The linear correlation coefficient between $f$ and $f_\mathit{cut}$, for the 8 measurements $f_\mathit{cut}\le f_\mathit{ce}$ observed during events A, B and C, is $c=0.96$.}
\label{fig2}
\end{figure*}


\end{article}

\end{document}